\documentclass[final,5p,times,twocolumn,authoryear]{elsarticle}
\usepackage{float}
\usepackage{longtable}
\usepackage{amsmath}
\usepackage{relsize}
\usepackage[authoryear]{natbib}
\usepackage[normalem]{ulem}
\usepackage{hyperref}
\usepackage{subcaption}
\usepackage{booktabs}
\usepackage{xcolor}
\usepackage{multicol}
\usepackage{graphicx}
\def\tsc#1{\csdef{#1}{\textsc{\lowercase{#1}}\xspace}}
\tsc{WGM}
\tsc{QE}
\tsc{EP}
\tsc{PMS}
\tsc{BEC}
\tsc{DE}

\usepackage{amssymb}
\usepackage{lipsum}

\begin{document}
\let\WriteBookmarks\relax
\def\floatpagepagefraction{1}
\def\textpagefraction{.001}

\begin{frontmatter}



\title{Luminosity-Dependent Variations in the Secondary Maximum of Type Ia Supernovae and Their Connection to Host Galaxy Morphology}

   \author[label1]{Jagriti Gaba}
\affiliation[label1]{organization={Department of Basic and Applied Sciences, School of Engineering and Sciences},
            addressline={GD Goenka University},
             city={Gurugram},
             postcode={122103},
             state={Haryana},
             country={India}}

 \author[label3]{Rahul Kumar Thakur} 
\affiliation[label3]{organization={School of Computer Science and Engineering, IILM University},
             city={Gurugram},
             postcode={122011},
             state={Haryana},
             country={India}}

\author[label1]{Dinkar Verma} 
\author[label1]{Naresh Sharma}

\author[label1]{Shashikant Gupta\corref{cor1}}
\cortext[cor1]{Corresponding author
}

\begin{abstract}
Type Ia supernovae (SNe Ia) are considered standardizable candles and are therefore important probes of the universe's expansion history and cosmic distances. In comparison to the optical and IR photometric observations, NIR light curves of SNe Ia are more uniform and are less affected by dust extinction; hence, they can provide more precise distance estimates. This study examines the relationship between the luminosity-dependent behavior of the  NIR secondary maximum ($t_2$) and the decline rate parameter ($\Delta m_{15}$) in the B Band. We analyzed 54 SNe Ia using linear, piecewise linear regression, and non-linear models along with non-parametric statistical techniques to examine the correlation between $t_2$ and $\Delta m_{15}$. Our results show that the secondary maximum timing varies among SNe Ia but exhibits a luminosity-dependent structure, with significant differences between SNe hosted in late and early-type galaxies. Two separate groups belonging to different host morphologies have been identified through our analysis, one containing brighter SNe and the other containing fainter SNe. These findings have important implications for improving the calibration of SNe Ia for cosmological applications.
\end{abstract}


\end{frontmatter}

\section{Introduction}
\label{introduction}

Type Ia supernovae (SNe Ia) are thermonuclear explosions in carbon-oxygen (CO) white dwarfs (WDs) in binary systems. They are triggered when the white dwarf approaches the Chandrasekhar mass via accretion or merger \citep{toonen2012supernova, wang2018mass, ruiter2025type}. SNe Ia are among the crucial probes in cosmology due to their high luminosities and relatively uniform peak brightness, which allow their calibration as ``standardizable candles" for measuring extragalactic distances \citep{khetan2021new, moresco2022unveiling, perivolaropoulos2024hubble,  hayes2025characterising}. Their importance was underscored by the discovery of accelerated expansion of the Universe, which relied on precise SN Ia distance measurements \citep{perlmutter1998discovery, riess1998observational,riess2004type, riess2004identification}. 

SNe Ia calibration for measuring cosmic distances involves empirical relationships among different observables, such as the Phillips relation \citep{phillips1993absolute}, which states that brighter SNe decline more slowly while dimmer ones decline faster. Modern light curve (LC) fitters such as SALT2 \citep{guy2007salt2}
or SNooPy \citep{burns2010carnegie} utilize multi-band photometric data for calibration. Through these calibration schemes, SNe Ia provide precise distance moduli across a wide range of redshift, enabling the construction of the Hubble diagram and the determination of key cosmological parameters. 
Unfortunately, optical LCs are heavily affected by dust, leading to variations in the reddening law and introducing significant uncertainties in measurement \citep{truong2022modeling}. Near-infrared (NIR) light curves of SNe Ia offer several major advantages that help to mitigate key calibration issues found in optical analyses \citep{BENGOCHEA20115}. Optical LCs require corrections using light curve shape and color, but residual scatter remains. However, NIR peak magnitudes of SNe Ia are intrinsically more uniform, even without shape or color corrections. Thus, NIR light curves reduce systematics from dust, empirical corrections, and host galaxy effects, making SNe Ia more reliable distance indicators \citep{wood2008type}. They significantly improve standardization and provide valuable physical insight into the diversity of SNe Ia. Because longer wavelengths are less affected by interstellar dust extinction, the NIR LCs of SNe Ia are more robust luminosity indicators \citep{Krisciunas_2017}.  

The NIR LCs of SNe Ia exhibit a prominent secondary maximum in the $J$, $H$, $K$, and $Y$ bands \citep{kasen2006secondary, dhawan2016reddening} and in $r$ and $i$ bands \citep{deckers2024ztf}. It arises approximately 20–40 days after the $B$-band maximum. This feature is understood to originate from the recombination of doubly ionized to singly ionized iron-group elements, i.e., FeIII to FeII and CoIII to CoII. The recombination takes place when the ejecta has cooled down to $\sim 7000$K and results in a redistribution of flux from the optical to the infrared as the ejecta cools and expands \citep{kasen2006secondary}. 

The timing of the second maximum $t_2$ in NIR LC is found to have an anticorrelation with the B band decline rate ($\Delta m_{15}$). It indicates that the supernovae that rapidly fall after maximum brightness (higher $\Delta m_{15}$) typically have shorter $t_2$, whereas slow decliners (small $\Delta m_{15}$) have longer $t_2$ \citep{kasen2006secondary, dhawan2015near}. 
 The decline rate of B band LC, $\Delta m_{15}$, is small in more luminous SNe Ia as the amount of $^{56}$Ni in such SNe is large, which can power the LC for a longer duration. On the other hand, fainter SNe Ia decline fast, i.e.,  $\Delta m_{15}$ is relatively larger for such SNe due to the smaller amount of $^{56}$Ni synthesized in the explosion \citep{10.1093/mnras/stad2851}. It has been observed \citep{dhawan2015near} that slower declining SNe Ia tend to exhibit a later and stronger secondary maximum, while fast-declining SNe tend to show an earlier and weaker secondary maximum. 

The anticorrelation between $t_2$ and $\Delta m_{15}$ can be used to estimate the peak luminosity and, hence, can serve as an alternative standardization tool \citep{dhawan2015near, dhawan2016reddening, gaba2025accurate}. It improves distance modulus estimates and decreases dispersion in the Hubble diagram \citep{dhawan2018measuring}. Recent studies also indicate that progenitor age, metallicity, and local environment may play important role in shaping the IR light curve structure \citep{10.1093/mnras/stv757}. The time of secondary maximum thus provides a valuable diagnostic of ejecta ionization, nickel mass, and progenitor system properties in SNe Ia.

The key factors affecting the timing and strength of secondary maxima may include opacity, the amount of iron-group elements, and the explosion mechanism \citep{deckers2024ztf}. 
The exact role of opacity evolution, ionization transitions from FeIII to FeII, and energy deposition remains quantitatively disentangled from other effects, such as density structure and mixing. Some peculiar events (e.g., 91bg-like, 02cx-like) deviate strongly from the relation. The reason for this deviation is not known. Host galaxy morphology, mass, and stellar population age correlate with $\Delta m_{15}$, so indirectly they may also influence $t_2$. We wish to investigate the following: (i) Does the same linear relation hold for both bright and faint SNe Ia, and (ii) Is there any effect of host galaxy morphology on the relation? 

The structure of this article is as follows. We discuss the methods and data used for analysis in Section 2. The results of our study are presented in Section 3, and we conclude in Section 4. 

\section{Data and Methodology}
We aim to explore the dependence of the second NIR maximum timing on Optical LC and host galaxy environment. In the following section, We discuss the photometry data of SNe Ia in both NIR and Optical bands used in the study. We also outline the statistical techniques employed in our study.

\subsection{Data}
\label{sec:data}
To study the anticorrelation between the NIR second maximum timing and the B-band decline rate, we require a sample of well-observed SNe Ia in both the NIR and B band. Although optical and NIR photometric observations of SNe Ia are well reported, not all SNe Ia show the second maxima in the NIR band, which is the main focus of the present work. We have taken 54 SNe from the Carnegie Supernova Project (CSP), which show prominent second maxima in the NIR band. CSP is a five-year study at the Las Campanas Observatory with the aim to get high-quality light curves of about 100 low-redshift SNe Ia in a well-defined photometric system \citep{contreras2010carnegie, burns2010carnegie, stritzinger2011carnegie, phillips2012near, burns2014carnegie}. Second maximum in $J$ band wavelength range ($1.1 \mu {m} > \lambda > 1.4 \mu {m} $), and the decline rate $\Delta m_{15}$ along with the tag of each SN are also available in \citep{dhawan2015near, dhawan2016reddening}. 

To further investigate the dependence of the relation on the host environment, we compile the morphology $T$ of the host galaxies of 54 SNe Ia (presented in the~\ref{sec:appendix}) from the SAI Supernovae catalog. The Sternberg Astronomical Institute (SAI) Catalog of Supernovae, which is a database of observations of supernovae, including SNe Ia \citep{tsvetkov2004_snicat}. 
\subsection{Methodology}
To ensure a comprehensive analysis of the data, a multi-faceted statistical approach is adopted, starting with the evaluation of both linear and non-linear patterns. The data analysis techniques used in this study are discussed below.

\subsubsection{Piece-wise Linear Regression and Non-linear Modeling}
\label{sec:Method-plr}
While a linear relationship between $\Delta_{m15}$ and $t_2$ has been previously reported, this correlation has not been explored extensively. In this study, we aim to study potential deviations from this linear behavior to uncover deeper insights into complex trends that a standard linear model might fail to detect. Consequently, we employ a dual approach using both piece-wise linear regression and a quadratic model. 

\textbf{Piece-wise Linear Regression:} By dividing the data set into segments and fitting each segment with its linear relation \citep{bemporad2021piecewise}, it can model changes in linear trends in different regions. The points where these segments meet, the breakpoints, allow for a more accurate relation between $t_2$ and $\Delta_{m 15}$. The following equation describes the piecewise linear regression with a single breakpoint. 
\begin{equation}
y = \left\{
\begin{array}{ll}
     a_1x + b_1, \,  x \in [x_0, x_1] \\ 
     a_2x + b_2, \,  x \in [x_1, x_2] \, ,
\end{array}
    \right.
\end{equation}
where $x$ and $y$ represent $\Delta_{m 15}$ and $t_2$, respectively. Each segment ($a_ix + b_i$) represents linear regression in a specific interval [$x_{i-1}, x_i$], and the breakpoint defines the boundary between the linear segments \citep{168933,1705,unknown}. 

The key steps in applying piecewise linear regression are (i) segmenting the data based on prior knowledge and (ii) fitting separate linear equations to each segment, ensuring continuity at breakpoints to avoid artificial discontinuities. 
To ensure the accuracy of our piece-wise regression, we utilize the Chi-square ($\chi^2$) statistic to determine the optimal breakpoint. This metric quantifies the difference between the model's predicted values and the actual measurements while accounting for inherent measurement errors. A small $\chi^2$ value indicates a "good fit," signifying that the predicted values align closely with the observed data; conversely, a large $\chi^2$ indicates a "poor fit". For this analysis, we define $\chi^2$ as follows:
\begin{equation}
\chi^2(p) = \sum_{i=1}^{n}
\frac{( y_i - f(x_i, p) )^2}
{\sigma_{y,i}^{2} + \left( \frac{\partial f}{\partial x}\,\sigma_{x,i} \right)^{2}} \, .
\end{equation}

Where $f(x_i, p)$ is the model prediction,  $\sigma_{y,i}$ and $\sigma_{x,i}$ are the experimental uncertainties and $\dfrac{\partial f}{\partial x}$ is the gradient (slope) of the model which scales the $x$-error into the $y$-dimension.

\textbf{Non-linear Model:} 
A quadratic model provides a standard and statistically sound way to test for deviation from linearity that allows smooth curvature in the relationship between $x$ and $y$ without introducing abrupt structural breaks. We employ the following quadratic model as an alternative to the linear regression: 
\begin{equation}
y = a + bx + cx^{2}  . 
\label{eq:quadratic}
\end{equation}
The parameters $a, b$, and $c$ represent the intercept, first-order rate of change of $y$ with respect to $x$, and the curvature of the relationship, respectively. 

\subsubsection{Comparison of the Models}
We employ the Akaike Information Criterion (AIC) and Bayesian Information Criterion (BIC) to cross-validate the results obtained from the single linear model, quadratic model, and piecewise linear regression. 
\\
\textbf{AIC:} It compares different models by computing the AIC \citep{akaike2003new} defined as:
    \begin{equation}
    \text{AIC} = - 2\, \log(L) +  2k ,
    \end{equation}
where $k$ represents the number of model parameters, and $L$ is the maximum likelihood. We estimated likelihood through $\chi^2$ defined as $$\chi^2 = \sum_i \Bigg(\frac{t_{2,i}^{obs} - t_2^{pred}(\Delta m_{15,i};a,b)}{\sigma_i}\Bigg)^2 ,$$ where $\sigma_i$ represents the observational uncertainties in $t_2$. AIC takes into account both the complexity of the model (number of parameters, $k$) and the goodness-of-fit (likelihood, $L$). The model with the lowest AIC value is favored. 

For single linear and the non-linear models, we define $\text{AIC}_{\text{lin}} = 2k - 2\log(L_{\text{lin}})$ and $\text{AIC}_{\text{non-lin}} = 2k - 2\log(L_{\text{non-lin}})$, respectively. For the groups in piece-wise linear regression we define $\text{AIC}_{\text{Groups}} = 2k - 2 \log\ ( L_{\text{Group1}} \times L_{\text{Group2}})$ \citep{singh2025does}. $L_{\text{Group1}}$ and $L_{\text{Group2}}$ are the likelihoods of the linear models for the subgroups, respectively. A basic linear regression model has only two parameters: the intercept and the slope. In contrast, a piecewise linear model has two parameters for each segment, thus more parameters in total. 
Lower AIC values indicate a better-fitting model, as they penalize models with excessive complexity, thereby reducing the risk of overfitting. To quantify the difference between the two models, i and j, we define 
\begin{equation}
\Delta \text{AIC} = \text{AIC}_{\text{i}} - \text{AIC}_{\text{j}}.
\end{equation}
$\Delta \text{AIC} \ge 2$ suggests a statistically significant difference between models $i$ and $j$, favoring the model with the lower AIC. 
\\
\textbf{BIC:} It allows eliminating models that are too complicated to avoid overfitting and favors models with lower BIC values defined as: 
\begin{equation}
    \text{BIC} = - 2\, \log(L) +  k\, \log(n) ,
    \end{equation}
where $n$ is the sample size. BIC is preferred over AIC when the sample size is large. 
To quantify the difference between the models $i$ and $j$, we define 
\begin{equation}
\Delta \text{BIC} = \text{BIC}_{\text{i}} - \text{BIC}_{\text{j}}.
\end{equation}
$\lvert \Delta \text{BIC}\rvert \ge 6$ suggests a strong preference for the model with lower BIC \citep{raftery1995bayesian}. 
\subsubsection{Mann-Whitney U Test}
If the piecewise linear regression suggests a break in the linear relation between $t_2$ and $\Delta m_{15}$, the SNe Ia can be divided into different subgroups. If so, we would like to understand the possible differences in the properties of host galaxies, such as morphology ($T$). We first confirm the normality of $T$ using the Shapiro-Wilk test by computing a $W$ statistic as: 
\begin{equation}
W = \frac{\left( \sum_{i=1}^{n} a_i x_{i} \right)^2}{\sum_{i=1}^{n} (x_i - \bar{x})^2} \, \,  ,
\end{equation}
where $x_{i}$ are ordered sample values and $a_i$ are constants based on the expected values of order statistics from a normal distribution, and $\bar{x}$ is the sample mean \citep{Hanusz_Tarasinska_Zielinski_2016}. The values of $W$ close to $1$ suggest that the data is drawn from a normal distribution. The null and alternative hypotheses are: 
$H_{Null}$: the data follows a normal distribution, and 
$H_{A}$: the data does not follow a normal distribution.
If $p-$value is less than the chosen significance level (usually 0.05), the null hypothesis is rejected. The Shapiro-Wilk test statistic gives $W \approx 0.9100$, and $p \approx 0.0006$ for variable $T$. This means the null hypothesis is rejected and the data is not normally distributed. 

Since the morphology of the host galaxy ($T$) is an ordinal variable and not normally distributed, we need to use a rank sum test. Thus, instead of the students' t-test to compare the properties of different subgroups, we apply the Mann-Whitney U test, also known as the Wilcoxon rank-sum test. It determines whether the distributions of two samples are likely to be the same. It is often used when the data does not meet the assumptions of a t-test, such as normality \citep{article2}. 
The formulas for the Mann--Whitney $U$ test are:
\[
U_{1} = n_{1}n_{2} + \frac{n_{1}(n_{1}+1)}{2} - R_{1},
\qquad
U_{2} = n_{1}n_{2} + \frac{n_{2}(n_{2}+1)}{2} - R_{2}.
\]
where $n_1$, $n_2$ are the number of elements in the first and second groups, respectively. $R_{1}$ and $R_{2}$ are the sums of the ranks for each group. The minimum of $U_1$ and $U_2$, i.e., $\rm{min}(U_1, U_2)$ is compared against the critical value $U_{crit}$ obtained from the Mann-Whitney $U$ table. The null hypothesis, i.e., the two groups obey identical distributions, is rejected if $U_{crit}< U$ calculated and vice-versa. 

If $n_1$ and $n_2$ are large, the distribution of U approximates a normal distribution. For the Mann-Whitney U value, the smaller of the two values, $U_1$ and $U_2$, is used. Here, $\mu$ is defined as the mean. 
\begin{equation}
 \mu = \frac{n_1 n_2}{2},
\end{equation}
and 
\begin{equation}
\sigma^2 = \frac{n_1 n_2 (n_1 + n_2 + 1)}{12}  \,  .
\end{equation}
$\sigma$ is defined as the standard deviation. After the mean and standard deviation have been estimated, $z$ can be calculated.
\begin{equation}
z = \frac{U - \mu}{\sigma} \, .
\end{equation}
The case of $z<-1.96$ or $z>1.96$ corresponds to $p<0.05$ and leads to rejection of the null hypothesis. 

\section{Results and Discussion}
\label{sec:Results}
In this section, we present the results of our analysis, which includes piecewise linear regression, the non-parametric Mann-Whitney U test, and clustering analysis, for 54 SNe Ia.

\subsection{Piecewise Linear Regression Analysis and Non-linear Model}
\label{sec:Results-plr}
We begin by establishing the overall relationship between the bolometric decline rate, $\Delta m_{15}$, and the timing of the near-infrared secondary maximum, $t_2$, for our sample of 54 SNe Ia. A standard linear regression across the full dataset yields the relation:

\begin{equation}
t_2 = 53.282 - 21.529 \Delta m_{15}  \, .
\end{equation}

This negative slope confirms the known anticorrelation between these parameters.

\begin{figure}[htbp]
    \centering
    \includegraphics[width=\linewidth]{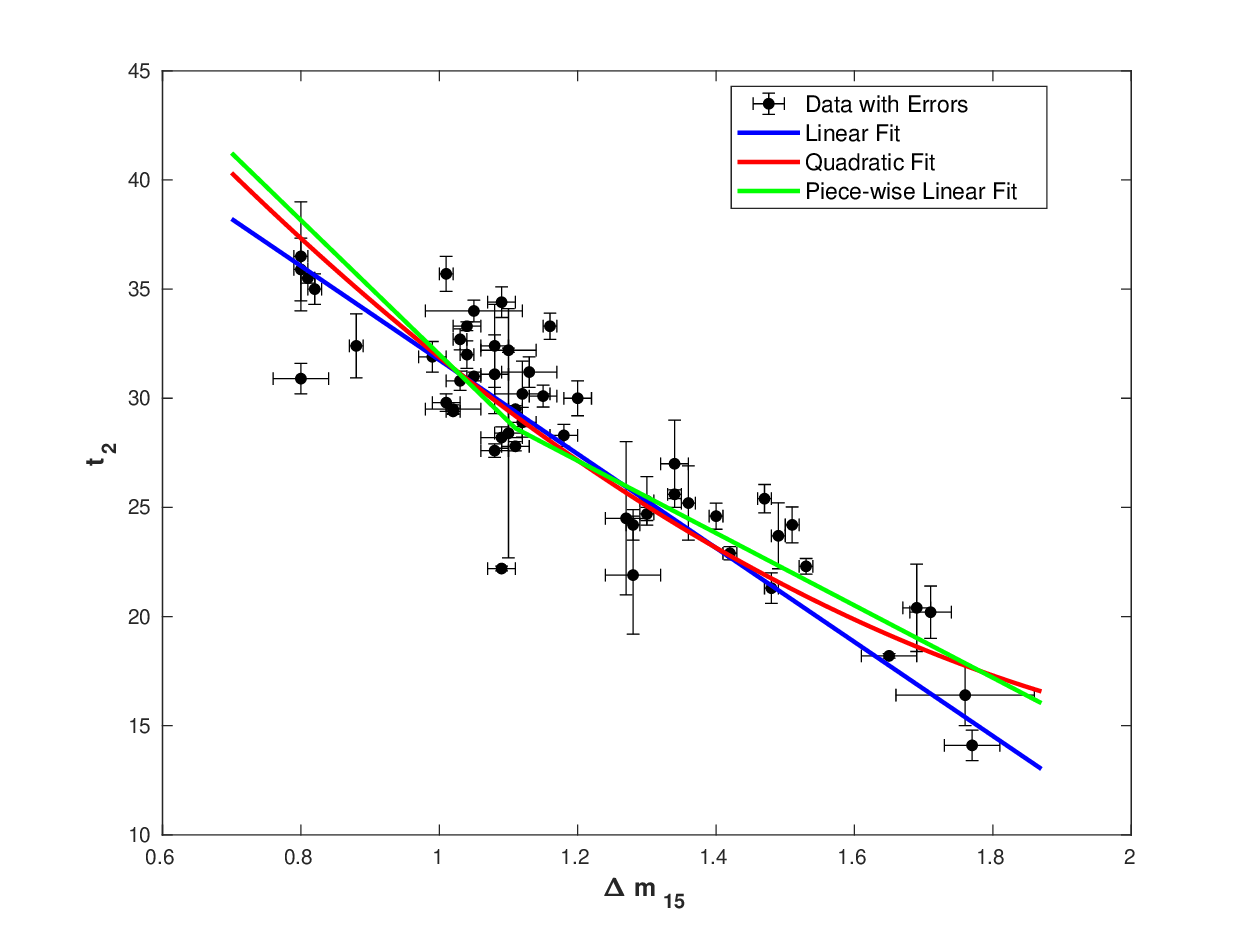}
    \caption{Comparison of linear, quadratic, and piecewise regression models fitted to the data.}
    \label{fig:regression_models}
\end{figure}

\textbf{Piece-wise Linear Regression:} To investigate a potential bifurcation in this relationship, we employed a piecewise linear regression model. This analysis identified a statistically significant breakpoint in the slope at $\Delta m_{15} = 1.11$ (Fig.~\ref{fig:regression_models}). The sample is thereby divided into two distinct groups:
\begingroup
\small
\begin{align}
\textbf{Group 1 (Slow Decliners; $\Delta m_{15} \leq 1.11$):}\quad 
& t_2 = 62.742 - 30.734\,\Delta m_{15}, \\
\textbf{Group 2 (Fast Decliners; $\Delta m_{15} > 1.11$):}\quad 
& t_2 = 47.015 - 16.565\,\Delta m_{15}.
\end{align}
\endgroup

\textbf{Non-linear Model:} 
In addition to the piece-wise approach, we fit a quadratic equation to the data to capture smoother non-linear variations. The resulting best-fit parameters are provided in Table~\ref{table:bestfit}, where the coefficient of the quadratic term remains small relative to the other terms. The specific relation is defined as:
\begin{equation}
   t_2 = 66.310 - 43.431 \Delta m_{15} + 9.002 \Delta m_{15}^{2} \, .
\end{equation}

\textbf{Model comparison:} 
To evaluate the efficacy of these different approaches, we compare the models using the Akaike Information Criterion (AIC) and the Bayesian Information Criterion (BIC). As shown in Table~\ref{table:aic_bic}, both AIC and BIC are highest for the standard linear model, underscoring its inadequacy compared to the models proposed in this study \citep{dhawan2016reddening}. Conversely, the piece-wise linear regression yields the lowest AIC and BIC values. These significant differences indicate that the piece-wise model is strongly favored over both the single linear fit and the quadratic model.

The change in slope (two groups in piece-wise linear regression) indicates that the physical link between the photometric decline rate and the ionization-state evolution of the ejecta is not uniform across all SNe Ia. The steeper slope for Group 1 suggests that $t_2$ varies more sensitively with $\Delta m_{15}$ for brighter, slower-declining events. In contrast, the shallower slope for the fainter, faster-declining events of Group 2 indicates a less sensitive relationship. Several potential factors may contribute to this bifurcation, including: (i) sensitivity to variations in the mass of $^{56}\text{Ni}$, (ii) the characteristics of the host environment, particularly its morphology, (iii) the propagation of radiation as influenced by ionization conditions, and (iv) systematic issues inherent in the observations \citep{kasen2006secondary, Kasen_2007, sullivan2010dependence}. In this study, we focus on the effect of host galaxy morphology by analyzing the relevant morphology data, which is discussed in the next section. 

\begin{table}[h!]
\centering
\caption{Best-fit values of various parameters for different models.}
\label{table:bestfit}
\begin{tabular}{@{}l r r r@{}}
\toprule
Model & a & b & c \\
\midrule
Single linear & 53.282 & -21.529 & - \\
Piece-wise $\Delta m_{15} \leq 1.11$ & 62.742 & -30.734 & -  \\
Piece-wise $\Delta m_{15} > 1.11$ & 47.015  & -16.565 & -  \\
Non-linear &     66.310 & -43.431 & 9.002 \\
\bottomrule
\end{tabular}
\end{table}
\begin{table}[h!]
\centering
\caption{AIC and BIC for a single linear fit (full data), a linear fit with a breakpoint, and a non-linear model. The smaller AIC and BIC favor the model with a breakpoint at $\Delta m_{15}=1.11$. }
\label{table:aic_bic}
\begin{tabular}{@{}l r r@{}}
\toprule
Model & AIC & BIC \\
\midrule
Single linear  & 733.740 & 737.718\\
Piece-wise linear  & 563.985 & 571.9410\\
Non-linear  & 682.336  & 688.303\\
\bottomrule
\end{tabular}
\end{table}
\subsection{Host Galaxy Morphology and the Mann-Whitney \textit{U} Test}
\label{sec:Results-mwut}

Visual inspection of the host morphology of the groups in Fig.~\ref{fig:regression_models} suggests an environmental separation, i.e., SNe Ia in Group 1 (slow decliners) appear predominantly associated with late-type galaxies (larger $T$ values), while those in Group 2 (fast decliners) are more common in early-type hosts (relatively smaller $T$ values). We quantified this morphological difference using the non-parametric Mann-Whitney \textit{U} test, as it is appropriate for the ordinal host morphology parameter (\textit{T}). The hypotheses for the test were defined as:
\begin{itemize}
    \item[$H_0$:] The host galaxy morphology distributions for both groups are identical.
    \item[$H_A$:] The host galaxy morphology distributions for both groups are different.
\end{itemize}
The test results are summarized in Table~\ref{tab:utest}, which rejects the null hypothesis at $95\%$ confidence level as the $p$ value is $< 0.05$. Thus, the test confirms that SNe Ia in the two photometrically distinct groups belong to host galaxies with fundamentally different morphological types.
\begin{table}[h!]
\centering
\caption{Results of the Mann-Whitney \textit{U} test for host galaxy morphology.}
\label{tab:utest}
\begin{tabular}{lc}
\toprule
Statistic & Value \\
\midrule
\textit{U} statistic & 244.5 \\
Expected $\mu$ & 364.5 \\
Standard deviation $\sigma$ & 57.80 \\
\textit{z}-score & $-2.08$ \\
\textit{p}-value & 0.038 \\
\bottomrule
\end{tabular}
\end{table}

The above finding is consistent with earlier studies, supporting the case of two distinct progenitor channels correlated with progenitor age and host environment \citep{10.1093/mnras/staa2940, burgazrefId0}. Brighter, slower-declining SNe Ia (Group 1) are preferentially found in late-type star-forming galaxies, suggesting a progenitor population with younger ages and shorter delay times \citep{Hamuy_2000, sullivan2010dependence}. In contrast, fainter, faster-declining SNe Ia (Group 2) are associated with older stellar populations in early-type, passive galaxies, indicative of older progenitors and longer delay times \citep{Hamuy_2000,sullivan2010dependence, pruzhinskaya2020dependence}. 

\section{Conclusion}
\label{sec:Conclusion}
In this study, we applied rigorous statistical techniques to examine the dependence of NIR second maximum timing on the B-band decline rate and host galaxy morphology for a sample of 54 SNe Ia. Our findings confirm that the timing of the NIR second maximum is anticorrelated with the B-band decline rate. Significantly, the piece-wise linear regression reveals a critical breakpoint at $\Delta_{m15} = 1.11$, suggesting the existence of two physically distinct subgroups within the SN Ia population. This division is further supported by Mann-Whitney U tests and Hierarchical clustering, which demonstrate that these subgroups are associated with systematically different host galaxy morphologies. These results align with recent studies and reinforce the hypothesis of different SN Ia populations belonging to different host environments \citep{10.1093/mnras/staa2940,larison2024environmental, burgazrefId0}. 

Our findings indicate important implications for the calibration of SNe Ia as cosmological distance indicators. This, in turn, may have potential consequences for cosmological parameter estimation and for tracing the cosmic expansion.

\section*{Acknowledgements}
Shashikant Gupta thanks SERB (India) for financial assistance (EMR/2017/003714).

\let\cleardoublepage\clearpage 
\appendix
\setcounter{table}{0}
\setcounter{figure}{0}
\section{Cluster Analysis Using Hierarchical Clustering Technique and Gower Distance}
\label{sec:clustering}
We employ hierarchical clustering, a machine learning technique, to confirm the groups identified by piecewise linear regression. Unlike flat clustering methods, it builds a hierarchy of clusters, typically visualized using a dendrogram, which shows the nested grouping of data and the distances at which clusters are merged. To handle mixed data, Gower distance is commonly employed, as it calculates pairwise dissimilarities by normalizing and merging several variable types. The distance between two data points is calculated by averaging normalized differences among all features. 
\begin{equation}
\quad D_{ij} = \frac{1}{p} \sum_{k=1}^{p} |x_{ik} - x_{jk}| = \frac{1}{p} \sum_{k=1}^{p} d_{ijk} \, ,
\end{equation}
where $p$ is the number of features, $d_{ijk}$ represents the partial distance between points $i$ and $j$ for feature $k$ \citep{article}. The average of the distances between all pairs of points, known as the average linkage method, was used for clustering \citep{emmendorfer2021generalized}. The Silhouette Score \citep{9260048}, which ranges from -1 to +1, indicates how well the data points fit within their cluster and how far they are from nearby clusters. It is calculated as:
\begin{equation}
s_i = \frac{b_i - a_i}{\max(a_i, b_i)} \,  ,
\end{equation}
where $a_i$ represents the inter-cluster distance and $b_i$ represents the nearest cluster distance. 

The silhouette score \citep{9260048} was used to determine the optimal number of clusters. The maximum silhouette score of $\approx 0.53$ was achieved for $k = 2$. 
 It suggests a two-cluster solution that yields the most well-separated and coherent grouping of the data.

\begin{figure*}[t]
  \centering
\includegraphics[width=0.95\linewidth]{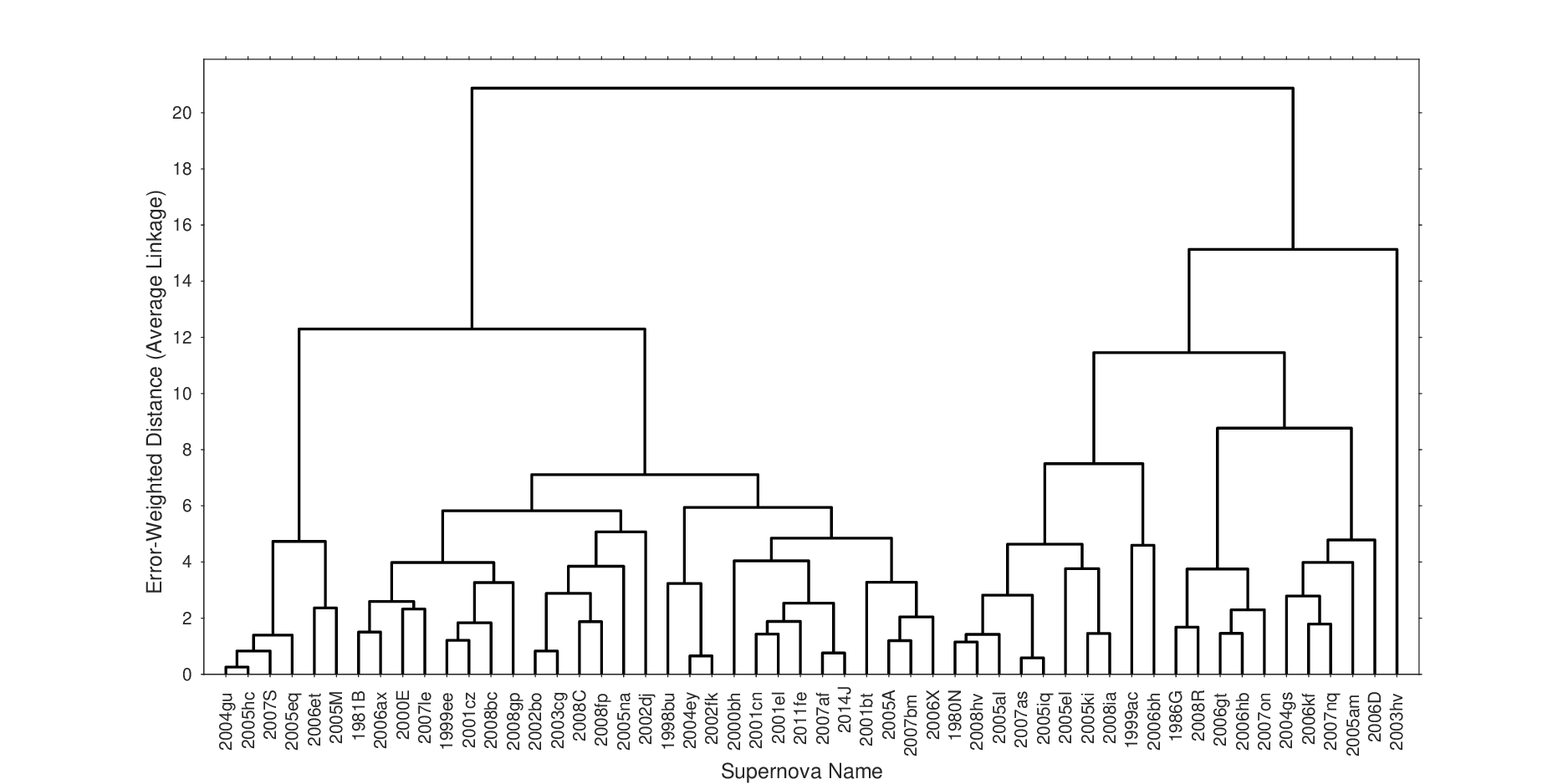}
     \caption{Hierarchical clustering dendrogram of Type Ia supernovae based on their photometric properties, constructed using the average linkage method. The vertical axis represents the distance or dissimilarity between clusters. The branching pattern reveals the hierarchical relationship among supernovae and suggests natural groupings based on similarities in their decline rates and secondary maximum characteristics.}
      \label{fig:dendogram}
\end{figure*}

The dendrogram in Fig.~\ref{fig:dendogram} resulting from the cluster analysis shows that two primary clusters bifurcate at a linkage distance of approximately 0.40. The characteristics of these clusters (see Table~\ref{tab:clusters} 
for details) are as follows:
\begin{itemize}
    \item[] \textbf{Cluster 1}: It contains 21 SNe Ia characterized by larger $\Delta m_{15}$ (i.e., faster decline) and smaller $t_2$ (i.e., earlier NIR secondary maximum).\\
    \item[] \textbf{Cluster 2}: It consists of 33 SNe Ia characterized by smaller $\Delta m_{15}$ (slower decline) and larger $t_2$ (later NIR secondary maximum).
\end{itemize}

\begin{table}[h!]
\centering
\caption{Characteristics of the two clusters identified through hierarchical clustering.}
\label{tab:clusters}
\begin{tabular}{lcc}
\toprule
Parameter & Cluster 1 & Cluster 2 \\
\midrule
Number of SNe Ia & 21 & 33 \\
Average $t_2$ (days) & $22.6$ & $31.5$ \\
Typical Host Morphology & Early-type & Late-type \\
\bottomrule
\end{tabular}
\end{table}

The two clusters obtained from the analysis agree with the piece-wise linear regression analysis. The convergence of these two distinct statistical methods provides robust and independent confirmation of two subpopulations within our SNe Ia sample. This substantiates the conclusion that the observed bimodality is inherent to the data and is not an artifact of the analysis techniques.


\section{SNe Ia Data}
\label{sec:appendix}
\setcounter{table}{0}
\setcounter{figure}{0}
\begin{table}[H]
\centering
\caption{Observed parameters of Type Ia supernovae (SNe), including the light-curve decline rate $\Delta m_{15}$ and its uncertainty, the secondary maximum time $t_2$ with its associated error, and the morphology parameter $T$ of 54 SNe Ia.}
\label{tab:full_sn_t2_T}
\begin{tabular}{lccccc}
\hline
SN & $\Delta m_{15}$ & $\sigma_{\Delta m_{15}}$ & $t_2$ & $\sigma_{t_2}$ & $T$ \\
\hline
1980N & 1.28 & 0.04 & 21.9 & 2.7  & -1.7  \\
1981B & 1.10 & 0.04 & 32.2 & 0.1  & 4.2   \\
1986G & 1.76 & 0.10 & 16.4 & 1.4  & -2.1  \\
1998bu& 1.01 & 0.02 & 29.8 & 0.4  & 1.8  \\
1999ac& 1.34 & 0.02 & 27.0 & 2.0  & 5.9   \\
1999ee& 1.09 & 0.02 & 34.4 & 0.7  & 4.0   \\
2000E & 0.99 & 0.02 & 31.9 & 0.7  & 3.9  \\
2000bh& 1.16 & 0.01 & 33.3 & 0.6  & 6.1   \\
2001bt& 1.18 & 0.02 & 28.3 & 0.5  & 3.9   \\
2001cn& 1.15 & 0.02 & 30.1 & 0.5  & 5.2   \\
2001cz& 1.05 & 0.07 & 34.0 & 0.5  & 4.8   \\
2001el& 1.13 & 0.04 & 31.2 & 0.7  & 5.9   \\
2002bo& 1.12 & 0.02 & 28.9 & 0.68 & 0.9   \\
2003cg& 1.12 & 0.04 & 30.2 & 1.5  & 1.1   \\
2003hv& 1.09 & 0.02 & 22.2 & 0.11 & -2.6  \\
2004ey& 1.02 & 0.01 & 29.4 & 0.14 & 4.3   \\
2004gs& 1.53 & 0.01 & 22.3 & 0.36 & -2.7  \\
2004gu& 0.80 & 0.01 & 35.9 & 1.44 & 1.5  \\
2005A & 1.08 & 0.02 & 27.6 & 0.31 & 4.9   \\
2005al& 1.30 & 0.01 & 25.3 & 1.11 & -3.3 \\
2005na& 1.03 & 0.01 & 32.7 & 0.48 & 1.0   \\
2006D & 1.47 & 0.01 & 25.4 & 0.65 & 2.0   \\
2006X & 1.09 & 0.03 & 28.2 & 0.49 & 4.0   \\
2006ax& 1.04 & 0.01 & 32.0 & 0.63 & 4.0   \\
2006et& 0.88 & 0.01 & 32.4 & 1.47 & 1.1   \\
2006gt& 1.71 & 0.03 & 20.2 & 1.2  & 4.5  \\
\end{tabular}
\end{table}
\begin{table}[H]
\ContinuedFloat
\begin{tabular}{cccccc}
\hline
SN & $\Delta m_{15}$ & $\sigma_{\Delta m_{15}}$ & $t_2$ & $\sigma_{t_2}$ & $T$ \\
\hline
2006hb& 1.69 & 0.02 & 20.4 & 2.0  & -2.6 \\
2006kf& 1.51 & 0.01 & 24.2 & 0.82 & -1.8  \\
2007S & 0.81 & 0.01 & 35.5 & 0.23 & 3.1  \\
2007af& 1.11 & 0.01 & 29.5 & 0.16 & 5.9  \\
2007as& 1.27 & 0.03 & 24.5 & 3.51 & 3.8  \\
2007bm& 1.11 & 0.02 & 27.8 & 0.21 & 5.0  \\
2007le& 1.03 & 0.02 & 30.8 & 0.44 & 4.9   \\
2007nq& 1.49 & 0.01 & 23.7 & 1.51 & -3.8  \\
2008C & 1.08 & 0.02 & 32.4 & 1.9  & 0.0   \\
2008fp& 1.05 & 0.01 & 31.0 & 0.2  & -1.6 \\
2014J & 1.10 & 0.02 & 28.4 & 5.71 & 8.0  \\
2002dj& 1.08 & 0.02 & 31.1 & 1.8  & -4.6 \\
2002fk& 1.02 & 0.04 & 29.5 & 0.2  & 3.9  \\
2005M & 0.80 & 0.04 & 30.9 & 0.7  & 3.5  \\
2005am& 1.48 & 0.01 & 21.3 & 0.7  & 1.1   \\
2005el& 1.40 & 0.01 & 24.6 & 0.6  & -1.9 \\
2005eq& 0.82 & 0.01 & 35.0 & 0.7  & 1.0  \\
2005hc& 0.80 & 0.01 & 36.5 & 2.5  & 2.4  \\
2005iq& 1.28 & 0.01 & 24.2 & 0.7  & 2.5  \\
2005ki& 1.36 & 0.01 & 25.2 & 1.7  & -2.7 \\
2006bh& 1.42 & 0.01 & 22.9 & 0.3  & 3.5  \\
2007on& 1.65 & 0.04 & 18.2 & 0.1  & -4.7 \\
2008R & 1.77 & 0.04 & 14.1 & 0.7  & -2.9\\
2008bc& 1.04 & 0.02 & 33.3 & 0.2  & 4.1   \\
2008gp& 1.01 & 0.01 & 35.7 & 0.8  & 1.5  \\
2008hv& 1.30 & 0.01 & 24.7 & 0.3  & -1.9  \\
2008ia& 1.34 & 0.01 & 25.6 & 0.2  & -2.2 \\
2011fe& 1.20 & 0.02 & 30.0 & 0.8  & 5.9   \\
\hline
\end{tabular}
\end{table}


\bibliographystyle{elsarticle-harv} 
\bibliography{cas-refs}






\end{document}